\documentclass[a4paper,11pt]{article}

\usepackage{xcolor}
\usepackage{microtype}
\usepackage{jinstpub} 
\usepackage{lineno}
\usepackage{hyperref}

\title{Novel Opaque Scintillator for Neutrino Detection}

\author[]{C. Buck,}
\author[]{B. Gramlich}
\author[]{and S. Schoppmann}
\affiliation[]{Max-Planck-Institut f\"ur Kernphysik,\\
  Saupfercheckweg 1, 69117 Heidelberg, Germany}

\emailAdd{christian.buck@mpi-hd.mpg.de}

\abstract{There is rising interest in organic scintillators with low scattering length for future neutrino detectors. Therefore, a new scintillator system was developed based on admixtures of paraffin wax in linear alkyl benzene. The transparency and viscosity of this gel-like material can be tuned by temperature adjustment. Whereas it is a colorless transparent liquid at temperatures around 40$^\circ$C, it has a milky wax structure below $20^\circ$C. The production and properties of such a scintillator as well as its advantages compared to transparent liquids are described.}

\begin{document}

\maketitle
\flushbottom

\section{Introduction}
\label{sec:intro}

From the very beginning, the technology of many detectors in the field of neutrino physics was based on transparent organic scintillators. Recently, a new concept of detector instrumentation was proposed (LiquidO) for which the scattering length in the scintillator has to be reduced to the cm level and below~\cite{LiquidO}. Light is collected in opaque scintillator material close to the interaction point of the ionizing radiation using optical fibers. This technology offers higher spatial resolution, better capabilities for particle identification and reduced constraints on the absorption properties as compared to standard liquid scintillator (LS) detectors.       

In our approach, this is achieved by the addition of paraffin wax to the well-known scintillator solvent linear alkyl benzene (LAB). We named this scintillator NoWaSH (New opaque Wax Scintillator, Heidelberg). Whereas at higher temperatures this mixture has optical properties similar to those of pure LAB, at lower temperatures the scintillator behaves more like solid wax. Besides the effect of a strongly reduced scattering length, the liquid gets highly viscous at and below room temperature. This has the advantage that the NoWaSH is rather insensitive to small detector leaks. Moreover, unwanted movement of impurities or additional components loaded to the NoWaSH by convection and precipitation effects are effectively reduced ensuring a highly stable system. Nevertheless, since it is produced and filled in the liquid state one still keeps the advantages of LS such as homogeneity and volume flexibility.  

One of these NoWaSH formulations with 10\,wt.\% wax concentration was tested in a small LiquidO prototype detector~\cite{LiquidO}. In this setup, 1\,MeV electrons were injected into a cell containing the scintillator. It could be demonstrated that the light collected at fibers within less than a few centimeters increases by more than a factor of 2 with the opaque NoWaSH instead of a transparent LS. At the same time, the signals further away are strongly reduced. This proves that the goal of local light confinement is achieved without significant loss of the total photoelectron signal. 

In this article we discuss details of the formulation, technical aspects and optical characterization of this new kind of scintillating material. The next section describes the chemical design of the NoWaSH, its composition and how it is produced. In section 3, the most relevant properties are discussed. The last section deals with metal loading capabilities of this system.

\section{Scintillator Design and Production}
\label{sec:production}
The scintillator is a three component system of LAB, paraffin wax and PPO (2,5-diphenyloxazole). All these components are known for low absorption of near UV light, high radiopurity, suitable safety properties, chemical inertness and moderate costs. The good performance of a LAB and PPO combination was demonstrated and studied in several multi-ton scale neutrino projects~\cite{DC, DB, RENO, Stereo, SNO, JUNO}. The paraffin wax is chemically similar to the n-paraffins also used in the LS of different neutrino detectors~\cite{Kamland, DC}.   

As solvent and main component in the NoWaSH test samples, LAB (CAS no.: 67774-74-7) from Helm AG (Helmalab D) was chosen. It was taken from the same batch used to produce the \textsc{Stereo} scintillators~\cite{StereoLS} and characterized accordingly. The pastillated white paraffin wax (CAS no.: 8002-74-2) used for the studies described in this article has a melting point between 52--54$^\circ$C (Fisher Chemical P/0600/90). It consists of long hydrocarbon chains with more than 20 carbon atoms per molecule. The paraffin wax for the proof-of-principle study in LiquidO~\cite{LiquidO} was from a different supplier (Aldrich 327204, melting point 53--58$^\circ$C). The wavelength shifter PPO (CAS no.: 92-71-7) is added to provide an emission spectrum peaking around 370\,nm. 

In the NoWaSH production, the paraffin wax pellets are added to the LAB in a containment which is heated to about 60$^\circ$C, slightly above the melting point of the wax. Then, the PPO is added at a concentration of 0.3~wt.\% and the mixture is stirred until all components are dissolved. As the scintillator cools down to $20^\circ$C it starts to solidify and obtains its soft wax structure. The growing wax crystals form a three-dimensional structure, with entrapped liquid hydrocarbons, leading to gelation. Three formulations were tested with different paraffin wax concentrations: 10~wt.\% (NoWaSH-10), 15~wt.\% (NoWaSH-15) and 20~wt.\% (NoWaSH-20). Details of the temperature dependent behavior of those samples are described in section~\ref{sec:temp}. The robustness of the structure was confirmed by loading a sample to a centrifuge. There, it was exposed for 5 minutes to a rotational speed of 4000\,rpm without any evident transformation despite the mechanical stress and high forces applied to the sample. This test indicates long term stability of the material under standard conditions.  

The largest samples produced so far were 10\,liter batches. Those could be used in an upscaled prototype LiquidO detector for more detailed studies after first measurements using the small prototype at the 100\,mL scale. The scalability of the production technology to larger masses seems to be feasible without major obstacles. 

\section{Properties}
\label{sec:properties}

\subsection{General}
In many laboratories, the choice of scintillator components is limited by safety regulations. Compared to other organic liquids, LAB is one of the least hazardous. The addition of paraffin wax and PPO does not change the hazard classification and even increases safety. Due to the rather solid structure, spills are no longer a risk and the vapor pressure of the odorless scintillator is very low. The LAB flash point of $\sim140^\circ$C also further rises with an increasing amount of the paraffin wax. 

The density of the produced NoWaSH samples is between 0.84 and 0.85\,g/mL at $20^\circ$C with volume expansion towards higher temperatures (see section~\ref{sec:temp}). The refractive index of the scintillator was determined to be 1.48 using a Kr\"uss DR201-95 refractometer. This value is similar to the one of pure LAB. During detector filling, the containments and filling lines should be heated to $40^\circ$C. At this temperature, the kinematic viscosity of the NoWaSH-20 is 5.6\,mm$^2$/s only slightly above the one of pure LAB.

\begin{table}[ht]
\caption[Radiopurity]{Upper limits on activities from different radioisotopes in the paraffin wax (90\% C.L.).}
\label{radio}
\begin{center}
\begin{tabular}{|l|ccccccc|}
\hline
Isotope & U-238 & Ra-226 & Th-228 & Ra-228 & Cs-137 & Co-60 & K-40\\
\hline
Activity [mBq/kg] & $<86.5$ & $<2.17$ & $<2.79$ & $<1.66$ & $<0.85$ & $<0.40$ & $<5.26$\\ 
\hline
\end{tabular}
\end{center}
\end{table}

Since this type of scintillator may have application in experiments requiring low backgrounds, radiopurity of the components is important. The radiopurity of LAB and PPO is already known from others projects. The LAB-based LS in Daya Bay achieved radiopurities for uranium (U) and thorium (Th) of $<10^{-12}$\,g/g and JUNO is expecting to achieve concentrations in the $10^{-15}$--$10^{-17}$\,g/g range~\cite{Lombardi:2019epz}. The radiopurity of liquid n-paraffins with shorter chains was also demonstrated~\cite{Kamland, DC}. A 2\,kg sample of the paraffin wax was monitored in the underground laboratory of our institute using the HPGe spectrometer GIOVE~\cite{GIOVE}. No $\gamma$-line signal above the background of the detector could be identified. Upper limits on typical radioimpurities derived from this measurement are listed in table~\ref{radio}. 

The properties related to light production and propagation are essential in a scintillator detector and will be discussed next. 

\subsection{Optical Properties}
\begin{figure}[htbp]
\centering 
\includegraphics[width=0.55\textwidth]{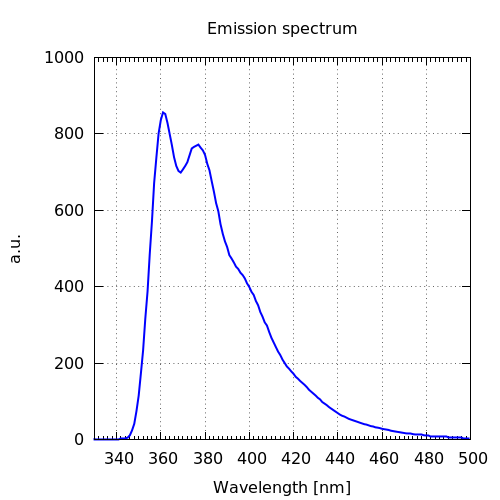}
\caption{\label{fig:1} The scintillator emission spectrum plotted here was determined for a sample with $\sim80$\,wt.\% LAB, $\sim20$\,wt.\% wax and a PPO concentration of 0.3\,wt.\%.}
\end{figure}

Scintillators based on LAB typically produce 50\% of the photons of anthracene for the same electron energy deposition~\cite{LABLY}. This translates into about 9000\,photons per MeV. The addition of 10--20\% of optically inactive paraffin is not expected to reduce the primary light yield (LY) by more than 5\%~\cite{CPL}. Direct measurements of the NoWaSH LY and comparisons to transparent LS are difficult due to the low scattering length. In a cell of 1\,cm length the measured LY of the NoWaSH-20 was found to be $>$80\% when compared to a transparent scintillator with pure LAB as solvent at similar PPO concentrations. The emission spectrum of a NoWaSH-20 sample is shown in figure~\ref{fig:1} as measured in a Varian CaryEclipse fluorimeter using a triangular 1\,cm cell. The light emission is mainly in the wavelength region between 350 and 400\,nm.

\begin{figure}[htbp]
\centering 
\includegraphics[width=0.55\textwidth]{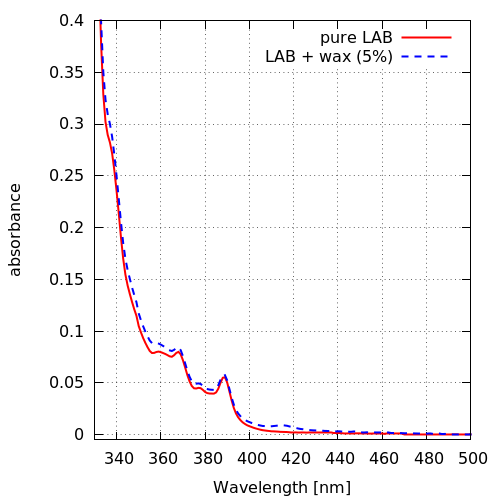}
\caption{\label{fig:2} The wavelength dependent absorbance is shown for pure LAB and a LAB/wax mixture with 5\,wt.\% paraffin. The samples were measured in a 10\,cm cell.}
\end{figure}

The attenuation length of pure LAB at 430\,nm was determined to be 23\,m~\cite{StereoLS}. At shorter wavelengths, in the region of the PPO emission below 400\,nm, the LAB absorption as well as self-absorption of PPO increases. However, the absorbed light in this region has a high probability to be re-emitted. The paraffin has no strong absorption bands in the region of interest as demonstrated in figure~\ref{fig:2}. In this figure, the absorbance of a pure LAB sample is compared to a mixture of wax/LAB at a ratio of 1:20. Both were measured in an UV/Vis photospectrometer (Agilent, Cary 4000) using a 10\,cm cell. In the wavelength region of interest, around the PPO emission peak, the structure of the absorption spectrum is clearly dominated by the LAB contribution. The high transparency of liquid n-paraffins and paraffin-based mineral oils in the optical and near UV region is known from several experiments~\cite{Kamland, DC}. There is a little bump around 420~nm from the paraffin wax. This is presumably from small impurities in the paraffin sample and might be different for other suppliers or wax batches. If the absorbance contribution from the paraffin is scaled to a scintillator system with 20\% wax fraction, the absorption length contribution of this component would be above 2~m in the region from 370--410\,nm. This is negligible as compared to the scattering length which is below 1\,cm. The light below 400\,nm is mainly absorbed by the fluorescent PPO or LAB and re-emitted with high probability.      

\begin{figure}[htbp]
\centering 
\includegraphics[width=0.55\textwidth]{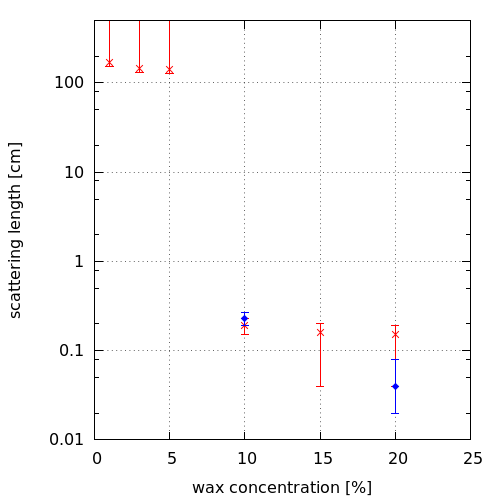}
\caption{\label{fig:3} The scattering length at 400\,nm is estimated for different wax concentrations from absorbance measurements in a 1\,cm cell (red points). The temperature during the measurements was $(23\pm1)^\circ$C. In the transparent regime, below 5\% wax concentration, the scattering length is more than 1\,m. At 10\%, crystallization is setting in and it drops below 2\,mm. Further increasing the wax content up to 20\% decreases the scattering length. The blue points illustrate the results of cross-check measurements performed in a 2\,mm cell.}
\end{figure}

The absorbance of several paraffin wax/LAB combinations was measured at room temperature ($23^\circ$C) inside a 1\,cm cell as shown in figure~\ref{fig:3}. For wax concentrations from 1--5~wt.\% the samples are still transparent. Since both scattering and absorption can contribute to the attenuation of the light beam going through the medium, the first three values given in figure~\ref{fig:3} should be interpreted as lower limits for the scattering length. The scattering length is well above 1\,m up to a wax concentration of 5\,wt.\%. At 10\,wt.\% and above, the measured absorbance increases by orders of magnitude. As shown above, this is not due to real absorption of the components. Since the probability of a photon scattering inside the cell to be detected in the photospectrometer is small, we interpret the measured attenuation of the light beam as scattering length. The observed scattering length is less than 2\,mm for wax concentrations above 10\,wt.\%. The given uncertainties include effects from temperature variations and a possible bias from photons scattered back into the beam. For the mixtures with more than 10\,wt.\% wax, the result was cross-checked by an independent measurement in a cell with a path length of only 2\,mm. For the NoWaSH-10 sample, a consistent value for the scattering length was found supporting the result obtained with the 1\,cm cell. For higher concentrations, the values in the 2\,mm cell were below the 1\,cm cell indicating contributions from scattered photons in the detected light beam, as mentioned above. The values obtained for the scattering length fulfill the requirements of a detector aiming for light confinement and spatial resolution on the 1\,cm level.

\subsection{Thermal Characteristics}
\label{sec:temp}
\begin{figure}[htbp]
\centering 
\includegraphics[width=0.7\textwidth]{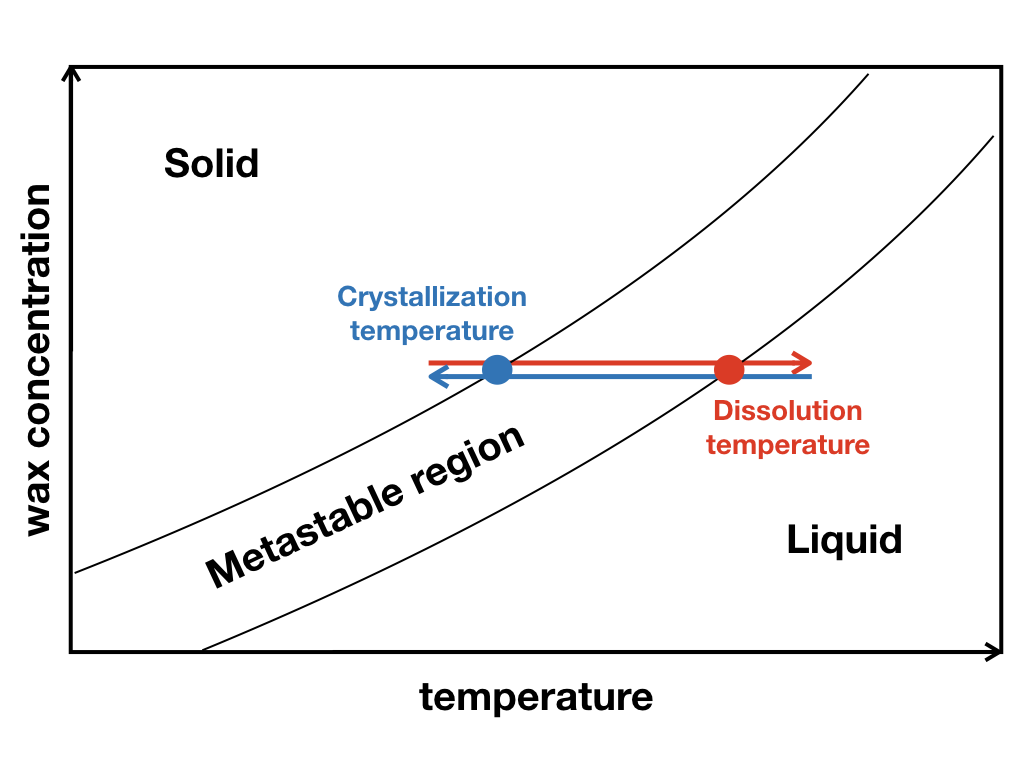}
\caption{\label{fig:4} This schematic illustration shows the three regions for an organic/liquid wax mixture: a liquid phase for high temperatures/low wax concentrations, a solid phase for low temperatures/high wax concentrations and a metastable region in between. Furthermore, the crystallization temperature at which crystals form while cooling the liquid phase (blue arrow), as well as the dissolution temperature at which the wax fully dissolves while heating the solid phase (red arrow), are indicated. The WAT is close to the dissolution temperature.}
\end{figure}

The transition between the liquid and solid phase of high molecular weight paraffins mixed with oil is widely discussed in literature. Such mixtures are created in crude oil production and transportation when n-paraffins precipitate out of solution and start to crystallize at cold inner surfaces of pipelines. In the field of petrochemistry, the solid-liquid thermodynamic equilibrium temperature is sometimes called the Wax Appearance Temperature (WAT) or cloud point~\cite{Andr}. When the liquid solution at high temperature is cooled down, the formation of first wax crystals typically starts below this WAT at the crystallization temperature. In between these temperatures, there is a metastable temperature region of supersaturation as illustrated in figure~\ref{fig:4}, for which the width depends on the cooling rate. The faster the temperature change, the wider is the metastable region. A linear dependence of the crystallization temperature and the cooling rate was reported~\cite{Andr}.

In addition, there is a correlation between the cooling rate and the crystal structure in such waxy oils. At faster cooling rates more crystals with smaller average size are formed. Moreover, a more regular and uniform distribution was reported for fast cooling~\cite{grow, Visi}. For slow cooling rates, the crystalline structure is expected to be more stable. Crystals start as small and thin platelets. They grow and form chains of crystals which become interconnected. The crystal sizes start with less than 5\,$\upmu$m slightly below the WAT and increase to typical values of approximately 20\,$\upmu$m at low temperatures~\cite{grow}. Parameters other than the temperature history influencing the crystal growth are pressure and composition. For example, it was suggested that higher fractions of iso-paraffin in the wax tend to favor microcrystalline or amorphous solids as compared to pure n-paraffins~\cite{Ronn}.

\begin{figure}[htbp]
\centering 
\includegraphics[width=0.7\textwidth]{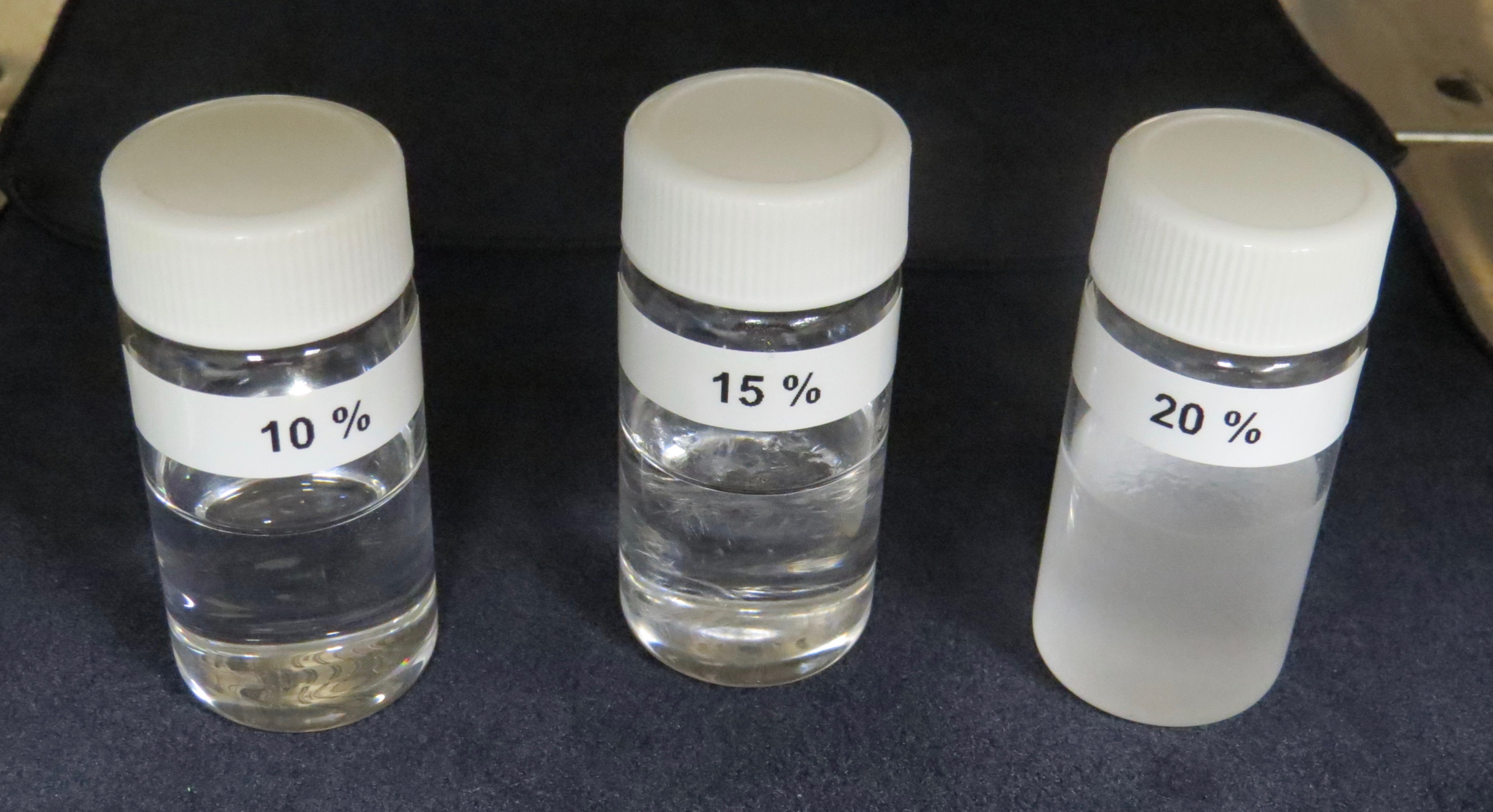}
\caption{\label{fig:5} The picture shows three samples with 10\% (left vial), 15\% (middle vial) and 20\% (right vial) of paraffin wax at $30^\circ$C. The scintillator was cooled from above $40^\circ$C at a cooling rate of about 2\,K/hour and kept overnight at $30^\circ$C. The outer diameter of the vials is 2.8\,cm. At $25^\circ$C all the samples are opaque.}
\end{figure}

After production at elevated temperatures, the NoWaSH samples start as transparent colorless liquids. As the temperature decreases they lose transparency and slowly solidify due to crystallization of the paraffin wax. In figure~\ref{fig:5}, three samples with different paraffin concentrations are shown at a temperature of $30^\circ$C. Whereas the NoWaSH-10 is still transparent, the NoWaSH-20 is already opaque for distances of more than 1~cm. The NoWaSH-15 is still rather transparent, but the formation of first crystals is already visible. At 25$^\circ$C all samples look similar to the NoWaSH-20 at $30^\circ$C. However, there is a difference in the rigidity at this temperature. The NoWaSH-20 is below, the other two samples still above the pour point. If the cooling rate is too slow, the crystals forming first might precipitate on the bottom of the vessel resulting in an inhomogeneous distribution of the n-paraffin in the scintillator. The Aldrich paraffin wax with slightly higher melting point also shows a slightly higher cloud point as compared to the samples using the Fisher Chemical wax.

For the case of the NoWaSH-20 (wax from Fisher Chemical) we observe crystallization around 32--33$^\circ$C when slowly cooling down the scintillator. Coming from low temperature and slowly heating up the mixture, we find full dissolution at 35--36$^\circ$C. In between, around 34$^\circ$C, there is the metastable region. At increasing temperatures, we observe a cloudy gel in this region (red arrow in figure~\ref{fig:4}) whereas there is still a transparent liquid, when coming from the high temperature side (blue arrow in figure~\ref{fig:4}). The difference of about 3$^\circ$C is only slightly higher than the extrapolated 1.6$^\circ$C width between the crystallization and dissolution temperature for a 20\% paraffin wax in mineral oil mixture at an infinitely slow temperature change~\cite{Andr}. For lower wax concentrations, the WAT decreases and the metastable region width increases. The WAT and metastable region width can also be affected by the PPO or impurities as well as mechanical stirring. 

A detector using a wax-based scintillator would probably be filled at elevated temperatures around 40$^\circ$C for which the medium can still be pumped through a piping and filter system. Assuming detector operation is at room temperature, there is a temperature difference between the filling and the data taking period of more than 10$^\circ$C. The change in the volume of the medium is sizeable in this transition. Therefore the thermal expansion coefficient is a crucial parameter for the design of the detector and auxiliary systems.

\begin{figure}[htbp]
\centering 
\includegraphics[width=0.55\textwidth]{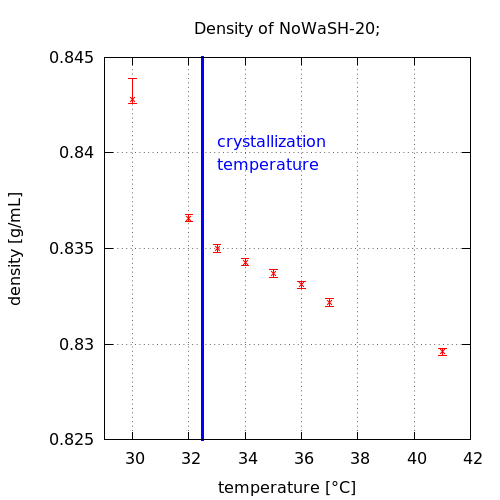}
\caption{\label{fig:6} The plot shows the temperature dependence of the scintillator density. The series of measurements started at 41$^\circ$C and then the scintillator was slowly cooled down. The error bars indicate the estimated uncorrelated uncertainty between data points. An extra systematic uncertainty contribution was added to the upper bar at 30$^\circ$C to account for the possible effects of gas bubbles, as described in the text.}
\end{figure}

In figure~\ref{fig:6}, the thermal expansion/compression is shown for a NoWaSH-20 sample. There is a linear behavior above the observed crystallization temperature. These data are suggesting a volumetric coefficient of thermal expansion around $7\cdot10^{-4}$\,K$^{-1}$ above 33$^\circ$C which is similar to other organic LS. A steeper increase of density is observed as the mixture cools down below the temperature of crystallization. The value at 30$^\circ$C plotted in figure~\ref{fig:6} was obtained after waiting at constant temperature for about 2 hours. The value could even be slightly underestimated in case of trapped gas bubbles. To avoid those, it is preferable to cool the scintillator in its containment from the bottom. This could be also relevant in large detectors, in particular if the diameter of the container decreases towards the top. 

\section{Boron Loading}
\label{sec:metal}
Many applications in the field of neutrino physics call for metal loading of the organic scintillator. Some examples are the gadolinium (Gd) loading to detect neutrons after the inverse beta decay of antineutrinos~\cite{DC, DB, RENO, Stereo}, boron (B) loading in active veto systems~\cite{Wright:2010bu}, tellurium (Te)/neodymium (Nd) loading for neutrinoless double beta decay experiments~\cite{Shimizu:2019irj} or indium (In) loading for solar neutrino detection~\cite{Raghavan:1976yc}. Metal loading in the NoWaSH approach can be similar as for other LS. Our approach would be to first load LAB, using standard techniques, and then add the PPO and paraffin wax. A basic requirement is that the loading technique can persist and is stable during the time when the mixture is heated up to 60$^\circ$C. 

As a NoWaSH derivative, first feasibility studies were performed with metal doping via a boronic acid. In the past, there were approaches using trimethylborate (TMB) to achieve B-concentrations of several percent~\cite{Green, Wang:1999mt}. Due to the toxicity of the TMB, we tested tributylborate (TBB, (C$_4$H$_9$O)$_3$B) instead, which is advantageous in terms of chemical hazard classification. The scintillator mixture contained about 60~wt.\% TBB, 20~wt.\% LAB, 20~wt.\% n-paraffin and a PPO-concentration of 0.8\,wt.\%. The TBB is a colorless liquid at room temperature and miscible with the LAB. However, it reacts rapidly with moisture and decomposes in water. Stability is improved by the solid structure of the scintillator provided by the paraffin wax. With a boron fraction of 4.7\% in TBB, the B-loading in the sample is 2.8\%. The most relevant isotope for neutron detection is $^{10}$B with a thermal neutron capture cross section of about 4000 barn. Without isotopic enrichment, the $^{10}$B-concentration (19.9\% natural abundance) is therefore 0.6\% in such a scintillator. The LY with such B-loading was determined to be ($60\pm6$)\% of the NoWaSH-20 corresponding to $\sim5000$~photons/MeV. 

Absorption measurements of TBB in cyclohexane have shown that the TBB molecule is highly transparent in the wavelength region of the scintillator emission. This implies the 40\% reduction in light response is not expected to be from absorption. Instead, this reduction can be explained by the factor 4 lower LAB fraction. This aromatic molecule is mainly responsible for the production of the scintillation light. However, there is a non-linear relation between the light yield and aromatic fraction~\cite{CPL}. The molar extinction coefficient of TBB at 300\,nm was determined to be 0.02\,L/(mol$\cdot$cm) to be compared to more than 30000\,L/(mol$\cdot$cm) for PPO at the same wavelength. Therefore, the quenching effect of TBB, even at the very high loading of 60~wt.\%, is expected to be moderate as confirmed in the measurement.

In general, the main challenge in the use of metal doped LS is to keep chemical and optical stability, in particular of the absorption length. Due to the tiny cross-section for neutrino interactions, the experiments in the field typically take data for several years to obtain reasonable statistics. Several experiments in the past using metal loaded LS were limited by the degradation of the attenuation length. One of the big advantages in the opaque scintillator scenario is that the constraints on this parameter are more relaxed as compared to transparent LS detectors. First, this opens the possibility of using techniques for high loadings which were rejected from attenuation length limitations so far. Second, a reduction of the absorption length would be less critical as long as it stays well above the scattering length. As demonstrated above, the scattering length in the NoWaSH samples is in the mm range whereas typical absorption lengths are on the meter level.

Since the scintillator system presented in this article is rather new, there are no long term stability data available yet. However, improved stability is expected due to the reduced mobility of molecules in the wax structure as compared to LS and the chemical inertness of the paraffin. At the same time, the crystallization helps to avoid precipitation of components. The centrifuge test described in section~\ref{sec:production} confirms the stability of the structure and that the wax will not easily separate from the LAB. First samples look stable and homogeneous by eye on a time scale of several months. Nevertheless, there are some caveats concerning this scintillator system. Operation of a detector in the metastable temperature region should be avoided. This could lead to a separation of wax and LAB. This implies temperature control is an important prerequisite. Concerning the boron scintillator described above, special attention has to be taken on the handling of the TBB. This boron molecule is highly reactive in contact with air and could potentially diminish the overall NoWaSH stability. Long term stability needs to be confirmed in future studies.

\section{Summary}
The idea of using an opaque medium as target in a large scale neutrino detector offers new possibilities for the scintillator design in such an experiment. The novel approach presented here offers encouraging characteristics for such an application. Admixtures of paraffin wax to suitable scintillator solvents as LAB result in scattering lengths in the mm range without significant absorption. Light yield losses are small and radiopurity estimates promising. A wax-like structure and the possibility to tune optical properties by temperature adjustment might provide additional benefits as compared to standard transparent liquid scintillators.    

\acknowledgments
The authors would like to thank Anatael Cabrera, LAL (Orsay, France), for triggering the idea on using opaque scintillators in neutrino detection and many fruitful discussions. Moreover we thank all people involved in the Micro-LiquidO measurement, in particular Christine Marquet and Michael S.~Pravikoff from CENBG (Bordeaux, France).

\end{document}